\documentstyle[12pt]{article}
\def\be{\begin{equation}}
\def\ee{\end{equation}}
\def\bea{\begin{eqnarray}}
\def\eea{\end{eqnarray}}
\newcommand{\bra}[1]{\langle #1 |}
\newcommand{\ket}[1]{| #1 \rangle}
\def\<{{\langle}}
\def\>{{\rangle}}
\def\vx{{\vec x}}
\def\vp{{\vec p}}
\def\vP{{\vec P}}
\def\vK{{\vec K}}
\def\sh{{\sf h}}

\def\F{{\cal F}}
\def\O{{\cal O}}
\def\P{{\cal P}}
\def\M{{\cal M}}

\date{ }
\begin{document}

\title{From Light Nuclei to Nuclear Matter\\ The Role of Relativity ?
\footnote {Dedicated to  the 80th Birthday of Hermann K\"ummel} }
\author{F. Coester\\
%\address
\small Physics Division,Argonne National Laboratory, Argonne, IL 60439, 
USA\\ \small E-mail: coester@anl.gov}

\maketitle

\abstract{The success of non-relativistic quantum dynamics 
in accounting for the binding energies and spectra  of light
nuclei with masses up to $A=10$ raises the question whether
the same dynamics applied to infinite nuclear matter agrees
with the empirical saturation properties of large nuclei.
The simple unambiguous relation between few-nucleon and many-nucleon 
Hamiltonians is directly related to the Galilean covariance of 
nonrelativistic dynamics. Relations between the irreducible unitary
representations of the Galilei and Poincar\'e groups indicate that
the ``nonrelativistic'' nuclear Hamiltonians may provide 
sufficiently accurate approximations to Poincar\'e invariant mass operators.
In relativistic nuclear dynamics based on suitable Lagrangeans
the intrinsic nucleon parity is an explicit, dynamically relevant, degree of
freedom and the emphasis is on properties of nuclear matter. The success 
of this approach suggests the question  how it might account for the
spectral properties of light nuclei. }

\section{Introduction}
The purpose of this paper is to examine some questions raised by the successes 
of ``non-relativistic'' dynamics of light nuclei\cite{carlson} on  one hand and  
``relativistic'' effective  theories of nuclear matter\cite{serot} on the other.  Can the
successful  few-nucleon Hamiltonian also describe  nuclear matter with comparable accuracy?
Is the success of ``nonrelativistic'' Hamiltonian dynamics compatible with the fundamental
requirement that space-time symmetry be realized by  unitary representations of the
Poincar\'e group\cite{wigner,haag_weinberg} on the Hilbert space of states?
Can many-body quantum mechanics formulated on tensor products of Dirac spinor functions
account for the successful features produced by Lagrangean
field theories?

\section{Standard Hamiltonian Dynamics}

Standard Hamiltonian nuclear dynamics is a
mathematically  consistent  framework for the
quantitative description of nuclei.  The form of
the two-nucleon Hamiltonian is suggested by
quantum field theory while the quantitative
details are adjusted to fit a vast  array of
available two-nucleon data. Much smaller
three-body forces are similarly determined.
Assuming that higher order multi-nucleon forces
are  negligible the Hamiltonian is then completely
determined for any nucleon number. Conserved
electromagnetic currents  consistent with these
nuclear forces are available.\cite{carlson}  Detailed
numerical  work has produced remarkable agreement
with many data. The computations produced the spectrum of the Galilei$\;$ 
invariant Casimir Hamiltonian
\be
{ \sh}:= \sum^A_{i=1}{\vp\,_i^2\over 2m} +\sum
V_{ij} +\sum V_{ijk} -{\vP\,^2\over 2Am}
\ee
with an accuracy of  $\sim 1\%$ for $A\leq
10$,\cite{carlson}  so that larger discrepancies with data
can be attributed to an inadequacy of the Hamiltonian.
The three-body potentials are essential for the agreement with the data.

This success raises  questions whether
the  Hamiltonian $H$ of $A$ nucleons
\be
H= \sum^A_{i=1}{\vp_i\,^2\over 2m}+\sum_{i<j<A} V_{ij}+ \sum_{i<j<k<A}V_{ijk}\; , 
\label{HAF}
 \ee
can also account for the low-energy properties of large nuclei.
The Hamiltonian (\ref{HAF}) defines the Fock-space Hamiltonian $H=H_0+V$ independent of the
nucleon number. This Hamiltonian is a number-conserving polynomial in the nucleon creation 
and annihilation operators, $c^\dagger(\vx)$ and $c(\vx)$, or their Fourier transforms
$c^\dagger(\vp)$ and $c(\vp)$. The Hamiltonian is Galilei covariant,
\be
[\vK,H_0]= i\vP\;, \qquad [\vK, V]=0\; ,\qquad [K_j,P_k]=im{\cal A}\delta_{jk}
\; ,
\ee
with 
\be
{\cal A}:=\int d^3 x\, c^\dagger(\vx)\, c(\vx)\; ,\qquad
\vP:=\int d^3 p\, c^\dagger(\vp)\,\vp\, c(\vp)\; ,
\ee
and the Galilean boost generator $\vK$ defined by
\be
\vK:=m \int d^3 x\, c^\dagger(\vx)\,\vx\, c(\vx)\; .
\ee
For a given Hamiltonian the saturation properties of homogeneous  nuclear 
matter present a mathematically well defined problem. 

The Fermi-gas ground state $\ket{\Phi}$ defined by
$a(\vx) \ket{\Phi}=0$, and $b(\vx) \ket{\Phi}=0$ with
\bea
a(\vx)&:=&{1\over (2\pi)^{3/2}}\int d^3 p\,\theta(|\vp|-k_F)c(\vp)e^{i\vp\cdot \vx}\cr\cr
b(\vx)&:=& {1\over (2\pi)^{3/2}}\int d^3 p\,\theta(k_F-|\vp|)
c^\dagger(\vp) e^{i\vp\cdot \vx}
\eea
is the cyclic vector of a Fock-space, $\F$, spanned by polynomials $\O$ of the creation operators
$a^\dagger(\vx)$ and $b^\dagger(\vx)$ applied to $\ket{\Phi}$ with the norm 
\be
\Vert \O\ket{ \Phi}\Vert^2= \bra{\Phi}\O^\dagger \O\ket {\Phi}^2
\ee
The  density $\rho_0$ is then
\be
\rho_0:=\bra{\Phi}c^\dagger(\vx)c(\vx)\ket{\Phi}= {4\over 
(2\pi)^3}\,{4\pi\over  3}\,k_F^3\; .
\ee

The ground state of homogeneous nuclear matter $\ket {\Psi}$ cannot be represented by a vector
in this Fock space. (Haag's theorem \cite{haag})  It can, however, be
defined by a linear functional $\<\O\Phi|\Psi\>$
over the Fock space  vectors $\O\ket{\Phi}\in \F$,\cite{fcoester} normalized to
$\<\Phi|\Psi\>=1$. This representation can be realized by $\ket{\Psi}= e^S\ket{\Phi}$ where $S$ is
a polynomial in the particle and hole creation operators, $a^\dagger$ and $b^\dagger$.
The operator $S$ is a sum of monomials, $S_n$, of  $n$ particle and $n$  hole creation operators.
Invariance under translations implies $S_1\equiv0$.
Any functional of this form is annihilated by the operators
\be
A(x):=a(x)-[a(x),S]\; , \qquad\mbox{and}\qquad B(x):=b(x)-[b(x),S]\;,
\ee
$A(x)|\Psi\> =0$ and $B(x)|\Psi\> =0$.  The operator $S$ is determined by 
the requirement that $\ket{\Psi}$ is stationary,
\be
[H,A(x)]\ket{\Psi}=0\; ,\qquad\mbox{and} \qquad [H,B(x)]\ket{\Psi}=0\; ,
\ee
or equivalently
\be
 \left[([H_0,S]+e^{-S}Ve^S),a(x)\right]\ket{\Phi}=0\; ,\quad\mbox{and} \quad
\left[([H_0,S]+e^{-S}Ve^S),b(x)\right]\ket{\Phi}=0\;.
\ee 
The energy per nucleon, $\epsilon$, is then determined by  $H(\vx)$ and 
the operators $S_2$ and $S_3$
\be
\epsilon :={1\over \rho_0}\,\bra{\Phi}e^{-S}H(\vx)e^S\ket{\Phi}=
{1\over \rho_0}\left(\bra{\Phi}H_0(\vx)\ket{\Phi} +
\bra{\Phi}[V(\vx),(S_2+S_3)]\ket{\Phi}\right)
\; .
\ee
With contemporary computing power accurate numerical
results  should become available. 

There is an infinity of unitary transformations of the 
 Hamiltonian which leave the $S$-matrix unchanged.\cite{ekstein}
Applied to Hamiltonians with only two-body potentials these
transformations  generate many-body potentials. The truncated Hamiltonians
with only the transformed two-body potentials have different spectral properties.
The infinite manifold of scattering
equivalent two-nucleon potentials is not sufficient to produce 
agreement with the empirical saturation properties of nuclear matter.\cite{coester,day} 
Corrections of the Brueckner approximations are needed  for accurate results.\cite{day} 
With the three-nucleon potential required by the spectral properties of light nuclei, agreement
or disagreement with empirical saturation properties of nuclear matter should give an indication
whether the successful few-nucleon dynamics can also account for the properties of heavy
nuclei. 

\section{Poincar\'e Compliance}

While the mathematical framework sketched here  exists for all energies it is 
applicable only well below meson production thresholds. With that
limitation there remains the question what  corrections are needed to
satisfy the  requirements
Poincar\'e symmetry.\cite{wigner,haag_weinberg}. Galilean
invariant nuclear dynamics can provide  good approximations  to results of Poincar\'e invariant
dynamics because the irreducible representations of the two groups are closely
related. The little groups, $SU(2)$  are the same. With a mass operator $\M$ of 
 $A$  nucleons of mass $m$ and the Hamiltonian $H_R$ defined by $H_R:=P^0-Am$,
the identity
\be
{{P^0}^2-A^2m^2\over 2 Am}={\M^2-A^2m^2\over 2Am} +{\vP\,^2\over 2Am}
=H_R\left( 1+{H_R\over 2Am}\right)\; ,
\ee
suggests the definition of the relativistic Casimir  Hamiltonian
\be
\sh_R:={\M^2-A^2m^2\over 2Am}\;.
\ee
The binding energies, $E_B$ of nuclei are related to the point spectrum of the
Casimir Hamiltonian $\sh_B$ by
\[
\sh_B= E_B \left(1+{E_B\over 2 Am}\right)\; .
\]
The ``relativistic correction'' implied by this relation is less than
$0.5\%$. The relation of the Casimir Hamiltonian $\sh_R$ to the Hamiltonian $H_R$
is 
approximately
the same as for Galilean dynamics as long as the spectral projectors of the
Hamiltonian 
are restricted to
values $E\ll 2Am$.  For two nucleons the Casimir Hamiltonian 
\be
\sh= {\M^2-4 m^2\over 4m}\equiv {k^2\over m} +V
\ee
is identical to the Galilean expression.
Common ``quasi-potential'' reductions\cite{lomon} of 
Lorentz invariant  Green functions produce resolvents of $\M^2$.

The Bakamjian-Thomas construction\cite{bak} of Poincar\'e generators as functions
of the mass operator and kinematic quantities does not satisfy
cluster separability\cite{cluster1}. But scattering equivalent generators
satisfying  cluster separability  can be constructed recursively
for any nucleon number $A$.\cite{cluster2} For the calculation of binding
energies and low-energy excitations the A-body forces generated in this manner
are expected to be negligible for $A>3$.

\section{Dirac Dynamics}

It can be advantageous
to realize symmetries by embedding the physical states in a larger
pseudo-Hilbert space. Quantum electro-dynamics is a well-known example.\cite{bogol}
The characteristic feature of Dirac quantum mechanics is the introduction
of four-component wave functions designed to permit linear representations
of the  Lorentz Group. The intrinsic parity is the additional
virtual degree of freedom introduced in the transition from the spin
representation to the spinor representation.
The Lorentz invariant inner product is indefinite, but the physical
one-particle states form a subspace with a positive invariant inner
product. In the presence of external fields the projection into the
physical states is a functional of the external fields. With many-body
quantum mechanics formulated on tensor products of spinor functions
the projection onto the physical subspace will, in general, be
interaction dependent. With the additional degree of freedom (intrinsic parity),
Slater determinants of modified single particle spinor functions can represent 
dynamical effects that would appear as
correlations in ordinary many-body wave functions.

For a single free nucleon the physical states are specified by the
projection
\be
\P_+:= {\vec \alpha\cdot \vp +\beta m+\omega(\vp)\over 2 \omega(\vp)}
\;,
\qquad \omega:=\sqrt{m^2+\vp\,^2}\,,
\ee
where $\vec \alpha=\gamma_5\vec \sigma$ and the intrinsic parity $\beta$ are represented by
the matrices
\be
 \vec \alpha=\pmatrix{0&\vec \sigma\cr\vec \sigma &0}\; ,
\qquad\beta= \pmatrix{1 &0\cr 0&-1}\; .
\ee
Thus any physical Fock space state $\ket{\Psi}$ of free nucleons satisfies
\be
\int d^3 p\, c^\dagger(\vp)\P_+(\vp) c(\vp)\ket{\Psi} =\ket{\Psi}
\ee
When the interactions do not commute with the projection $\P_+$ 
the subspace of physical states depends on the interactions. 
Projection onto the free-nucleon Hilbert space would produce
effective many-body potentials.

By definition the Dirac-Fermi-gas state $\ket{\Phi}$ is invariant under
translations, $\vP \ket{\Phi}=0$, and satisfies
 \bea
a(\vp)\, \ket{\Phi}=0\; ,\qquad 
 b(\vp)\,\ket{\Phi}=0\qquad \mbox{and}\qquad  d(\vp)\,\ket{\Phi}=0
\eea
with
\bea
 a(\vp)&:=&\theta(|\vp|-k_F)\P_+(\vp)c(\vp)\; ,\qquad b(\vp):= \theta(k_F-|\vp|)
\P_+(\vp)c^\dagger(\vp) \cr\cr
d(\vp)&:=& [1-\P_+(\vp)]c(\vp)\; .
\eea
It follows that
\be
\int d^3 p c^\dagger(\vp)\P_+(\vp) c(\vp)\ket{\Phi} =\ket{\Phi}\;.
\ee
In the absence of interactions the Fermi gas is stationary. 

The coupled cluster representation of homogeneous nuclear matter $\ket{\Psi}=e^S\ket{ \Phi}$ is similar
to the ``non-relativistic'' representation, except  for the additional virtual degrees of freedom
in the Fock space built on $\ket {\Phi}$. Now $S_1\equiv 0$ is no longer required by
translational invariance. However,
\be
\tilde a(\vp)e^{S_1}\ket{ \Phi}=\tilde b(\vp)e^{S_1}\ket{ \Phi}
=\tilde d(\vp)e^{S_1}\ket{ \Phi}=0\;,
\ee
with
\bea
&&\tilde a(\vp):=\theta(|\vp|-k_F)\tilde\P_+(\vp)c(\vp)\; ,\qquad
\tilde b(\vp):= \theta(k_F-|\vp|)\tilde\P_+(\vp)c^\dagger(\vp)\;,\cr\cr
&&\tilde d(\vp):= [1-\tilde \P_+(\vp)]c(\vp)\; ,
\eea
where the modified projection
\be
\tilde \P_+(\vp):={ \vec \alpha\cdot \vec p +\beta m^*+\omega^*\over  2\omega^*}
\;,\qquad \omega*:=\sqrt{{m^*}^2+\vp\,^2}\;,
\ee
is a functional of $S_1$ which depends on the interactions. With the modified Fermi gas
$\ket{\tilde \Phi}$ satisfying 
$
\tilde a(\vp)\ket{\tilde \Phi}=\tilde b(\vp)\ket{\tilde \Phi}
=\tilde d(\vp)\ket{\tilde \Phi}=0\;,
$
and $\Vert \tilde \Phi\Vert=1$ the ground state of nuclear matter can be represented by 
$\ket {\tilde \Psi}:=e^{\tilde S}\ket{\tilde \Phi}$ with $\tilde S_1=0$.
The approximation $\tilde S=0$ corresponds to the ``relativistic mean-field''
approximation. Since important dynamical features are already
incorporated in $\ket{\tilde \Phi}$, the coupled cluster correlations
$\tilde S$ may be considerably weaker than those required with
standard Hamiltonian dynamics.

\section{Conclusions}
``Realistic'' Hamiltonians with two- and three-nucleon potentials are consistent with a vast array
of two-nucleon data and with spectral properties of light nuclei. 
Accurate nuclear matter results for realistic Hamiltonians should become available. 
There is no firm theoretical basis for a prediction that there are ``realistic'' Hamiltonians
which fit the empirical nuclear matter properties without additional
many-body forces.

Relations between  Galilei and Poincar\'e representations indicate that Galilei covariant
Hamiltonian dynamics may adequately  approximate a scattering equivalent Poincar\'e covariant
Hamiltonian dynamics.

Hypothetical ``realistic'' Hamiltonians acting on tensor products of spinor functions
might account for both light nuclei and nuclear matter without a need for many-body
forces.

\section*{Acknowledgments}
This work  was supported in part by the Department of Energy, Nuclear
Physics Division, under contract W-31-109-ENG-38 .


\begin{thebibliography}{99}
\bibitem{carlson}J. Carlson and R. Schiavilla, Revs. Mod. Phys. {\bf 70},743 (1998);
\newline Steven C. Pieper and R.B. Wiringa, Ann. Rev. Nuc. Science {\bf 51}
(2001);\\ S.C. Pieper, private communication.
\bibitem{serot} Brian D. Serot and John Dirk Walecka, Int. J. of Modern Physics E, (2001).
\bibitem{wigner} E. P. Wigner, Ann Math. {\bf 40}, 141 (1939).
\bibitem{haag_weinberg} R. Haag, {\em Local Quantum Physics} (Springer Verlag
 1992) Chap. I.3 ;\newline
S. Weinberg {\em The Quantum Theory of Fields} (Cambr.Univ. Press 1995) Chap.2.
\bibitem{haag}R. Haag, {\em Local Quantum Physics} (Springer Verlag
(1992) p.57 .
\bibitem{fcoester} F. Coester, {\em Lectures in Theoretical Physics}
{\bf XI B}, 157 \\(Gordon and Breach 1969).
\bibitem{ekstein}H. Ekstein, Phys. Rev. {\bf 117}.1590 (1960);\\
R.J. Furnstahl, H.-W. Hammer, Negussie Tirfessa, Nuc. Phys.A {\bf 689}, 846 (2001);
ref. 5, p. 103.
\bibitem{coester} F. Coester, S. Cohen, B. Day and C.M Vincent, Phys. Rev.C 
{\bf 1}, 769 (1970).
\bibitem{day} B.D. Day, Comments Nucl. Part. Phys. {\bf 11}, 115 (1983).
\bibitem{lomon} M.H. Partovi and E.H. Lomon, Phys. Rev. {\bf D2} 1999 (1970);\\
A. Klein and T.-S. H. Lee, Phys. Rev. D {\bf 10}, 4308 (1974).
\bibitem{bak}B. Bakamjian and L. H. Thomas, Phys. Rev. {\bf 92} 1300
 (1953)\\ B.D. Keister and W.N. Polyzou,Adv. Nucl. Phys. {\bf 20},225 (1991).
\bibitem{cluster1} 
L. L. Foldy, Phys. Rev.  {\bf 122} 275 (1961);\newline
F. Coester, Helv. Phys. Acta {\bf 38} 7 (1965).
\bibitem{cluster2}
S. N. Sokolov, Dokl. Akad. Nauk. USSR {\bf 233} 575 (1977) and\newline
Teor. Mat Fiz. {\bf 36} 193 (1978);
\newline F. Coester and W. N Polyzou, Phys. Rev. D {\bf 26} 1348 (1982);
\newline W. H. Klink and W. N. Polyzou, Phys. Rev. C. {\bf 54} 1189 (1996)
\bibitem{bogol} N. N. Bogolubov, A.A. Logunov, A.I. Oksak , L.T. Todorov,
{\em General Principles of Quantum Field Theory} (Kluver Acad. Pub. 1987)
Chapter 10.


\end{thebibliography}
\end{document}